\documentclass[11pt,titlepage]{article}

\usepackage{placeins}
\usepackage{float}
\usepackage{multirow}

\usepackage{wrapfig}
\usepackage{jinstpub} 
\usepackage{amsmath}

\hypersetup{
  pdfnewwindow=true,
  colorlinks=true,
  linkcolor=blue,
  citecolor=blue,
  filecolor=blue,
  urlcolor=blue,
}

\date{\today}
\author{N.~Akchurin, C.~Cowden, J.~Damgov, A.~Hussain, and S.~Kunori}
\title{Perspectives on the Calibration of CNN Energy Reconstruction in
Highly Granular Calorimeters}

\affiliation{Texas Tech University, Department of Physics and Astronomy,\\
Advanced Particle Detector Laboratory \\ Lubbock, TX, 79409, USA}

\abstract{
We present a study which shows encouraging stability of the response linearity for a simulated high granularity calorimeter module reconstructed by a CNN model to miscalibration, bias, and noise effects.  Our results also show an intuitive, quantifiable relationship between
these factors and the calibration parameters.
We trained a CNN model to reconstruct energy in the calorimeter module using simulated single-pion events; we then observed the response of the model under various miscalibration, bias, and noise conditions that affected the model input.  From these data, we estimated linear response models to calibrate the CNN.  We also quantified the relationship between these factors and the calibration parameters by regression analysis.
}


\begin{document}

\maketitle

\section{Introduction}
%
Calibration of depth segmented, non-compensating calorimeters has always
challenged experimenters. Undeterred, some methods have been developed to calibrate such devices. However, another trend in calorimeter design pushes the number of readout channels to amounts which  warrant sophisticated reconstruction methods which seem incompatible with
previously established calibration methods.  We have started experimenting with applications of machine learning models - particularly convolutional neural networks or CNNs - to reconstruct energy from the many signal channels coming from such devices.  The increasing channel count reduces
occupancy which reduces the sample size available in any one channel for calibration; the
increasing channel count also stresses the computational resources
available for calibration activities.
For example, a Cu/Si test stand $1\times 1\times 3\;\mathrm{m}^3$
with a $50\times 50\times 75$ segmentation scheme contains $187,500$
cells.  If this scheme is extended to a barrel plus endcap configuration
at LHC detector size scales, the combined system contains $\sim65$
million channels.  Whatever the solution may be, methods to calibrate these devices must possess an efficiency in computation, data usage, and speed of data collection.

This development raises several concerns in relation to
calibrating and extracting physical measurements with
confidence.  The inherent non-linearity of CNNs, or any neural network,
forces several questions.  In this paper, we confine ourselves to matters
related specifically to calibration.
Primarily, we focus on studying a CNN's sensitivity to variations
of the input data in order to determine the feasibility of any sort of
calibration.

In general, we have two options to correct the CNN's reconstructed
energy.
We can adjust the calorimeter readout channel values, the
inputs to the model, which we shall call pre-model or we can adjust the predicted energy from the CNN reconstruction, the CNN output, which we shall call post-model.
Historically, experiments have employed linear reconstruction methods, simple sums of cell responses, resulting in simple multiplicative factors to calibrate.

A full-size, combined electromagnetic and hadronic high granularity calorimeter of LHC experimental proportions would  have upward of 10s of millions of cells.
Attempting to calibrate millions of channels via traditional techniques
seems like an unreasonably large task.
The degrees of freedom appear simply
too large for experimenters to determine the calibration parameters with any manageable sized calibration sample.  When the calorimeter is segmented longitudinally,
the rear cells will always suffer from low occupancy and low statistics.
Similarly, the physics generating signals at that depth is
different than in the front portion of the calorimeter (assuming we are
not calibrating with muons). 
As others have reported before~\cite{GANEL1998621}, this scenario makes for some unfortunate outcomes.  It makes matters worse for ourselves since we now have to deal with a non-linear energy reconstruction as well.

We studied the effect of miscalibration and noise on a CNN trained on a
Cu/Si module having a $2\times 2\times 4\;\mathrm{cm}^3$
segmentation. In a previous paper, we described the reconstruction performance of this design~\cite{akchurin2021use}.  The CNN we trained operates as a correction to the sum of energy of the calorimeter cells; the convolutional layers of the CNN consume the raw calorimeter cell energies and we inject the sum of cell energies at a later layer of the network.
We found this architecture gives us excellent performance on hadron initiated showers.
The data consist of single charged pions ranging up to 500
GeV simulated by the GEANT4~\cite{ALLISON2016186}.  
We modified the simulated cell energies by applying
Equation~\ref{eq:miscal} which multiplies the
simulated energy in cell $i$ of event $j$ by a dimensionless miscalibration factor
$C_i$, it adds a global bias $b$, and a stochastic noise
$\epsilon$ specific to the cell and event.
The bias $b$ and noise $\epsilon$ carry units of MeV to add to the base GEANT4 energy scale.
\begin{equation}
\tilde{E_{ij}} = C_i E_{ij} + b + \epsilon_{ij}
\label{eq:miscal}
\end{equation}
We also applied a zero suppression of $0.6\,\mathrm{MeV}$ to the adjusted energy values
which is approximately the energy deposited by a single MIP.
The miscalibration factors are drawn from
distributions centered around unity such that multiplication with
simulated energy results in adjustments of the detector cell's response.  Similarly,
noise is drawn from distributions centered at 0 on an event by event
basis such that adding it to the miscalibrated energy results in a
fluctuation around the nominal value.
We draw the miscalibration and noise values from the beta distribution with both parameters set to two, $B_{2,2}$, 
and then we transform the variables to achieve the desired
mean and scale of variation\footnote{The $B$-distribution with symmetric parameters greater than 1 has both the median and mode equal to $\frac{1}{2}$~\cite{numrec}.}. 
We adopted this approach because it sufficiently perturbs the CNN's inputs while mitigating the effects of potentially extreme events arising from long tails.  We have no expectation to derive quantitative results that transfer to other detectors or models, and this approach allows us to deduce transferable, qualitative conclusions based on our specific quantitative results.
Equations~\ref{eq:calib_dist} and~\ref{eq:noise_dist}
show the transformations of marginal miscalibration and noise
distributions where $s_c$ and $s_\epsilon$ represent
miscalibration and noise scale factors, respectively.
\begin{equation}
	C = s_c ( 2B_{2,2} - 1) + 1
\label{eq:calib_dist}
\end{equation}
\begin{equation}
	\epsilon = s_\epsilon (2B_{2,2} - 1)
\label{eq:noise_dist}
\end{equation}
These transformations give us distributions of random values in the range $[1-s_c,\,1+s_c]$ and $[-s_\epsilon,\,s_\epsilon]$ for miscalibration and noise, respectively.

The correlation of noise and calibration amongst cells results from
physical aspects of a detector design.  Lab bench calibration of
sub-modules, common electronics and read out boards for groups of channels,
upstream material, {\it etc}. lead to groups of read out channels showing
similar noise and response characteristics.  In order to draw random
noise values and miscalibration factors from a joint distribution that captures
the variable dependencies, we turn to vine copulae~\cite{aascopula, bedford2002}.
Modeling a joint distribution with a copula allows us to consider the univariate, marginal distributions separately from the variable inter-dependencies ({\it e.g.} correlation).
Vine copulas further help us to structure the variable inter-dependencies by only considering pairs of partial correlations.  We divided the $50\times 50\times 75$ test stand module into smaller sub-module sections; we chose a vine structure such that cells within a sub-module are correlated with one another while sub-modules remain independent.  We randomly generated correlation parameters for each sub-module.  This approach allows us to introduce correlation in a way that mimics common detector readout electronics architectures.

Table~\ref{tab:scan_params} lists the parameter values we
scanned in a grid to study the CNN sensitivity.
We applied the parameters to
modify the GEANT4 simulated cell energies prior to scoring the CNN.
For convenience, we will refer to the parameter set with the vector
$\boldsymbol\theta$ with calibration, noise, and bias scales as the
respective elements.
\begin{table}
\centering
\caption{The calibration, noise, and bias
parameters scanned in a $5\times 8\times 3$ grid.}
\label{tab:scan_params}
\begin{tabular}{|c|c|c|} \hline 
	Calibration Scale & Noise Scale [MeV] & Bias [MeV] \\ \hline
0. & 0. & 0. \\ 
0.15 & 0.1 & -0.15 \\
0.2 & 0.1 & 0.15 \\
0.25 & 0.3 & \\
0.3 & 0.4 & \\
 & 0.5 & \\
 & 1. & \\
 & 1.5 & \\
\hline
\end{tabular}
\end{table}

\section{Performance Dependencies}

\begin{figure}[H]
\centering
\includegraphics[width=0.75\textwidth]{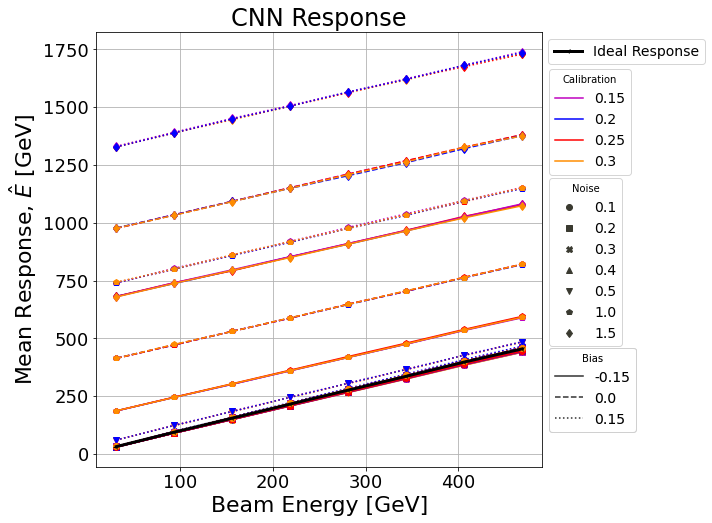}
\caption{The mean CNN response against initiating pion
energy for all of the scanned calibration parameters.  This collection
of curves gives some visual justification for our assumption of
linearity.
}
\label{fig:lin_assumption}
\end{figure}

We observe the mean CNN response remains linear with respect to the 
energy of the initiating particle.
Figure~\ref{fig:lin_assumption} shows the mean response as a function of the
initiating pion energy for
the calibration parameters $\boldsymbol\theta$ indicated in the legend.
As was discussed in~\cite{akchurin2021use}, the CNN acts like a correction to the simple energy sum.  This effect remains evident when the input distributions deviate from the training sample.
Since the zero suppression prevents negative contributions and the miscalibration factor is both multiplicative and positive, the energy sum is always positive.
We also notice in Figure~\ref{fig:lin_assumption} the significant over-estimation of energy by the CNN in some scenarios.  Typically, the sum of visible energy is a small fraction of the energy of the initiating particle.  Consequently, when the CNN amplifies the response of the raw signal with added noise and bias in its regression, the CNN responds with a significant over-estimation.  It is important to note, however, the observed linearity implies that we can simply re-calibrate the response.

\begin{figure}[H]
\centering
\includegraphics[width=0.9\textwidth]{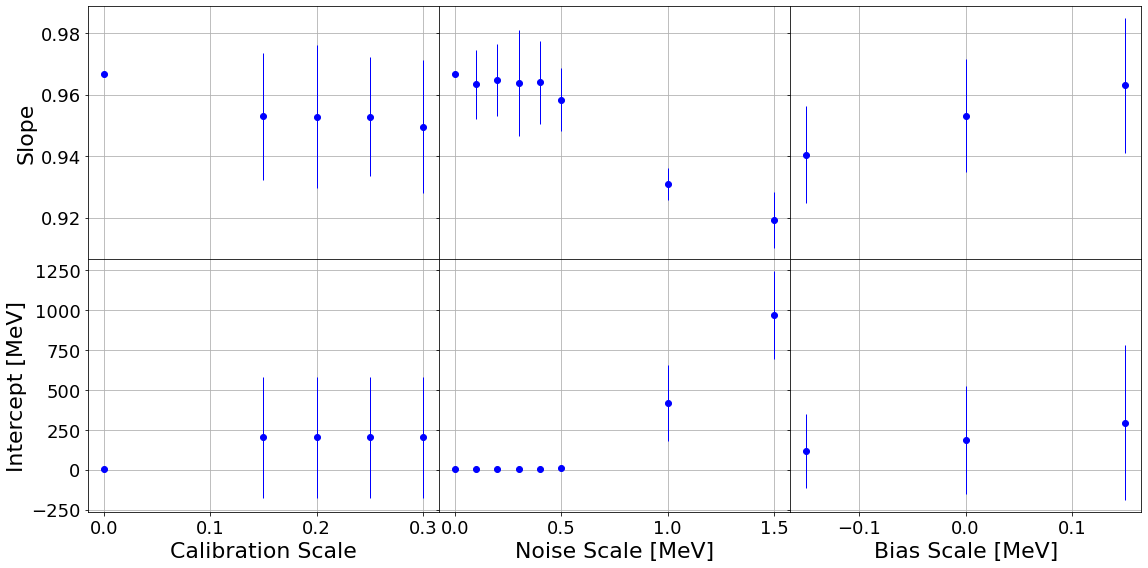}
\caption{Summary of response fits: the
subplots show the distributions of slope and intercept of the CNN
response against the calibration, noise, and bias scales.  The error bars represent the sample standard deviation.}
\label{fig:scan_summary}
\end{figure}

We fit linear models, $\hat{E} \sim a E_{beam} + b$, to the response data.  This
parameterization gives us the CNN's estimate's dependence
on the initiating pion energy.  One expects $a \sim 1$ and $b \sim 0
$.  Performing this fit for each point in the parameter scan reveals
the reconstruction's sensitivity to variations of the input data with
respect to the training data.  Figure~\ref{fig:scan_summary} summarizes
the results.
In Figure~\ref{fig:scan_summary}, we can see that noise and bias have a stronger
influence than the calibration scale.  We can also see the impact of the
zero suppression, $0.6\,\mathrm{MeV}$, in the relationship that noise scale has on the fit
parameters.

We then fitted independent linear models to the response slope, $a$, and
to the intercept, $b$.  We investigated fitting $a$ as dependent upon $b$,
but found it has a statistically insignificant contribution; therefore, we rejected the
dependence.  
Equations~\ref{eq:a} and~\ref{eq:b} enumerate the final estimated
models.

\begin{equation}
a = 0.963 - 0.299\cdot(s_\epsilon > 0.5)\cdot s_\epsilon -
	0.053\cdot(s_\epsilon > 0.5)\cdot s_b
+ 0.091\cdot s_b
\label{eq:a}
\end{equation}
\begin{equation}
b = -680\cdot(s_\epsilon > 0.5) + 1098\cdot(s_\epsilon >
	0.5)\times s_\epsilon + 2008\cdot(s_\epsilon > 0.5)\cdot
	s_b
\label{eq:b}
\end{equation}
We find a relatively weak relationship between the
bias and the slope compared to that of noise scale.
Interestingly, we observe an interaction between the noise scale and bias in
both fits. This implies that controlling noise should take the greater
share of concern in controlling accuracy.
These results are promising in that we see not only linear responses
throughout the scan but we also see linear relationships between the
scanned parameters and response parameters.
This result justifies a multiplicative calibration factor for post-model calibration; it also implies that we can interpret differences or changes of calibration constants in terms of relative differences or changes of properties such as noise scale.

\subsection{Energy Resolution Dependencies}
\label{ssec:resolution}
\begin{figure}[H]
    \centering
    \includegraphics[width=0.9\textwidth]{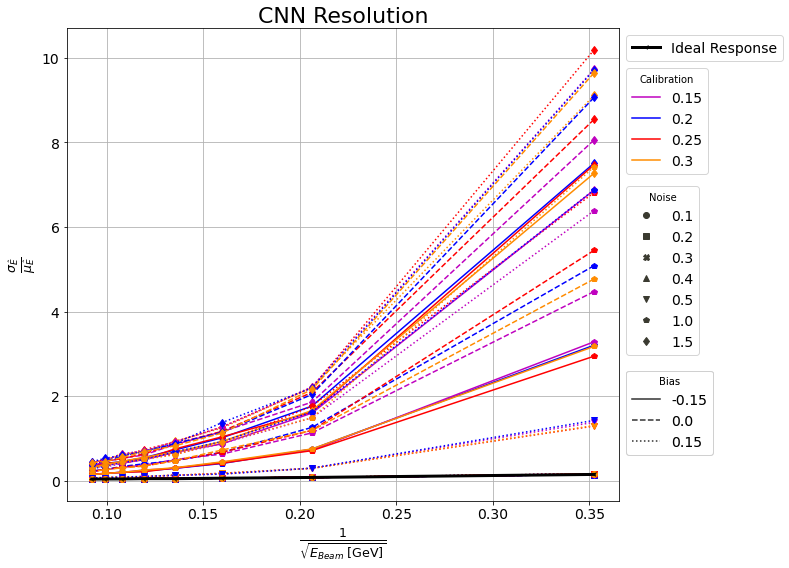}
    \caption{CNN reconstructed energy resolution {\it vs.} the inverse square-root of the beam energy for various miscalibration, noise, and bias parameters.}
    \label{fig:cnn_res}
\end{figure}

Similarly, we also studied the impact of miscalibration, noise, and bias on the  reconstructed energy resolution of the CNN.
This resolution study is based on a reduced beam energy scale, $[0, 125]\;\mathrm{GeV}$, to reduce the influence of leakage in higher energy events.
Figure~\ref{fig:cnn_res} shows the results of this scan in the traditional plot of resolution, $\frac{\sigma}{\mu}$, {\it vs.} $\frac{1}{\sqrt{E_{\rm beam}}}$.

\begin{figure}[H]
    \centering
    \includegraphics[width=0.9\textwidth]{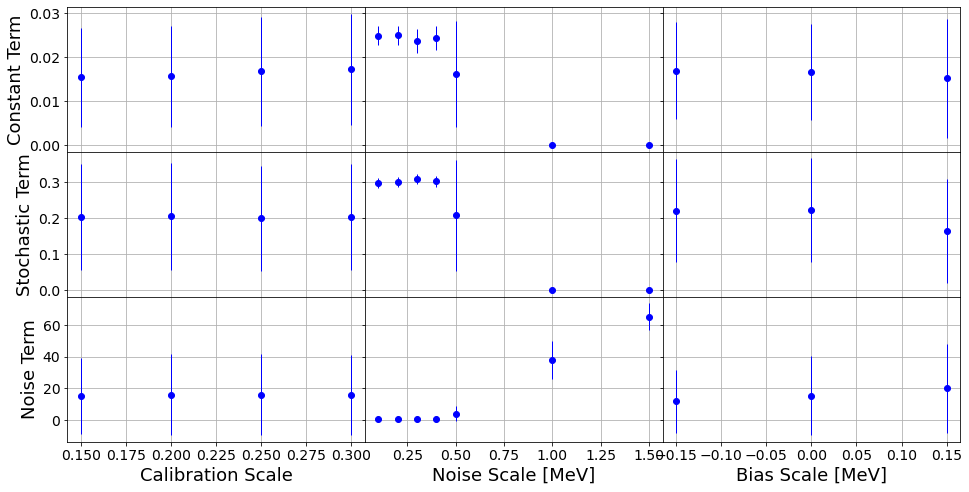}
    \caption{The estimated resolution fit parameters are summarized (mean and standard deviation) along the miscalibration, noise, and bias dimensions explored.}
    \label{fig:res_summary}
\end{figure}

We fit these resolution curves with a three term model,
\[
\frac{\sigma}{\mu} = \sqrt{\mathcal{C}^2 + \Big(\frac{\mathcal{S}}{\sqrt{E}}\Big)^2 + \Big(\frac{\mathcal{N}}{E}\Big)^2}
\;,
\]
with parameters corresponding to constant, stochastic, and noise terms respectively.
No variable selection or regularization is applied.  The parameters are estimated with the \texttt{lmfit}~\cite{lmfit} python package which implements the Levenberg-Marquardt least-squares methodology.  Figure~\ref{fig:res_summary} summarizes the estimated parameters.

In order to understand the CNN's energy resolution, we fit models to estimate the effects miscalibration, noise, and bias have on the constant, stochastic, and noise terms of the resolution parameterization.
As one can observe in Figure~\ref{fig:res_summary}, there is a distinct change of the parameter values at $s_\epsilon \sim 0.5$.  Further, by observing the distribution of residuals for a linear model fit to $s_c$, $s_\epsilon$, and $s_b$ alone, we can identify a contribution of bias to this distinct set of resolution behaviors.  Consequently, this motivates us to introduce a single variable to indicate the high noise and high bias region of simulation scenario space defined in Equation~\ref{eq:res_ind}.  The limited number of points along $s_b$ limits the precision with which we know when the effect truly turns on.  Nevertheless, we can understand this variable to indicate how noise and bias interact with the zero-suppression to allow non-signal effects to substantially contribute to the CNN resolution.
\begin{equation}
    I_{ND} = \Bigg\{ \begin{array}{c} s_\epsilon > 0.5;\;\mbox{or} \\  (s_b \geq 0.15)\, \&\, (s_\epsilon \geq 0.5) \\ \end{array}
    \label{eq:res_ind}
\end{equation}

Equations~\ref{eq:res_const},~\ref{eq:res_stoch}, and~\ref{eq:res_noise} report the final estimated effects.  We follow a step-wise backward selection process to remove insignificant variables from the fits.
\begin{equation}
    \log{\mathcal{C}} = -5.647 + 8.13 s_c - 28.696 I_{ND} s_c - 3.041 I_{ND} s_\epsilon - 7.722 I_{ND} s_b 
    \label{eq:res_const}
\end{equation}
\begin{equation}
    \log{\mathcal{S}} = -5.548 + 18.214 s_c - 36.076 I_{ND} s_c 	
    \label{eq:res_stoch}
\end{equation}
\begin{equation}
    \log{\mathcal{N}} = -0.887 s_c	+ 0.652 s_b + 4.762 	I_{ND} s_c + 2.355 	I_{ND} s_\epsilon + 1.123 	I_{ND} s_b
    \label{eq:res_noise}
\end{equation}
The log-transformation of the resolution parameters ensures non-negative estimated values.  We note that in the high noise and bias region ({\it e.g.} $I_{ND} = 1$) $\mathcal{C}$ and $\mathcal{S}$ are suppressed in favor of the $\mathcal{N}$ parameter.  We conclude from this that if noise and bias are not controlled below the zero-suppression, the energy resolution is no longer characterized by energy dependent Poisson statistics; it is rather characterized by an energy independent variance.  Conversely, in the low noise and bias region, we find $\mathcal{N}$ to be vanishingly small, and in this region, we find the resolution parameters determined mainly by the miscalibration scale.

\section{Discussion}
We find the consistent linearity of the CNN response under various input conditions encouraging for future use of these types of reconstruction models.
It is particularly noteworthy that the response linearity observed up to $500\; \mbox{GeV}$ extends well beyond the upper bound of the model build data that had an upper limit at $125\;\mbox{GeV}$; that is to say the CNN response linearity extrapolates beyond the build data.
The observed linearity implies that we can calibrate the estimated energy by multiplicative factor and additive bias offsets.  In fact, inverting the response fits discussed above achieves such a calibration.  With such an approach, one can calibrate a single test beam stand or improve observed response uniformity in a collider-based detector.
Although unexpected, a durable linearity of the CNN response may be a boon to future experiments.
Nevertheless, we should remain somewhat cautious because we have no guarantee that all CNNs will have the same linear response.

We also note that comprehending the impact of miscalibration, noise, and bias on the resolution parameters is more complicated.  Nevertheless, we can conclude from the results the relationships comport to our initial expectations.  When noise and bias are under control, the resolution follows the traditional constant and stochastic terms that depend on the miscalibration scale; however, when noise and bias are not under control, the resolution is characterized almost purely by the noise term.  The transition between low and high noise-bias regions may be better understood if we had modeled tails of the miscalibration and noise scales.  However, since the general trend contains no surprises, we do not think expending additional resources to study this would result in any interesting finding.  This point should be particularly salient when one considers that the particular numerical results do not translate form detector configuration to configuration.

These results should be read with the caveat that miscalibration, noise, and bias effects do not depend on depth.  Therefore, this study does not model well the effects of noise induced by long-term radiation damage.  Depth dependend effects may well lead to non-linear responses from the CNN.
We plan for future work to study how a CNN performs given real test beam data.

\acknowledgments{This work has been supported by the US Department of Energy, Office of Science
+(DE-SC0015592) and Texas Tech University, Office of the Vice President for Research and Innovation.}

\bibliography{calox.bib}

\end{document}